%% file: main_text.tex
\documentclass[aps,prd,twocolumn,showpacs,superscriptaddress,groupedaddress,floatfix]{revtex4}  

\usepackage{amsmath}
\usepackage{graphicx}  
\usepackage{dcolumn}   
\usepackage{bm}        
\usepackage{amssymb}   
\usepackage[mathlines]{lineno}
 \usepackage{epstopdf}
\usepackage{subfigure}
\usepackage{color}

\begin{document}

\widetext

\title{Probing Theories of Gravity with \\   Phase Space-Inferred Potentials of Galaxy Clusters}
\input author_list.tex       
\date{\today}

\begin{abstract}
Modified theories of gravity provide us with a unique opportunity to generate innovative tests of gravity. In Chameleon $f(R)$ gravity, the gravitational potential differs from the weak-field limit of general relativity (GR) in a mass dependent way. We develop a probe of gravity which compares high mass clusters, where Chameleon effects are weak, to low mass clusters, where the  effects can be strong. We utilize the escape velocity edges in the radius/velocity phase space to infer the gravitational potential profiles on scales of 0.3 - 1 virial radii. We show that the escape edges of low mass clusters are enhanced compared to GR, where the magnitude of the difference depends on the background field value $|\overline{f_{R0}}|$. We validate our probe using N-body simulations and simulated light cone galaxy data. For a DESI (Dark Energy Spectroscopic Instrument) Bright Galaxy Sample, including observational systematics, projection effects, and cosmic variance, our test can differentiate between GR and Chameleon $f(R)$ gravity models, $|\overline{f_{R0}}|= 4 \times 10^{-6}$ ($2 \times 10^{-6}$) at $>5\sigma$  ($>2\sigma$), more than an order of magnitude better than current cluster-scale constraints.\end{abstract}

\maketitle



\section{Introduction}

A variety of high precision observations all seem to indicate that our universe is currently undergoing accelerated expansion \cite{bschmidt,cosmicAccRev}. The most popular model that is consistent with these observations deploys the framework of General Relativity (GR) with an additional cosmological constant ($\Lambda$) that induces cosmic acceleration at late times. However, the key theoretical component of this concordant cosmological model, GR, is still poorly tested on megaparsec scales. This has given way for the proliferation of both novel ways of testing GR as well as gravitational theories that modify GR in cosmological scales (see \cite{Joyce,Koyama} for thorough reviews).

The main theoretical thrust behind models of modified gravity (MG) is that while we know GR and its weak-field limit (Newtonian gravity) work exquisitely well at the scales of binary pulsars and the solar system, we should be cautious when extrapolating out to much larger (i.e. cosmological) scales. For this reason, models of MG that successfully reproduce late time cosmic acceleration on large scales must also recover the predictions made by GR on small scales. 

To accomplish this, modified theories of gravity implement \textit{screening mechanisms} that attenuate the effect of  additional forces in high density regions. One such mechanism is the \textit{Chameleon mechanism} whereby the additional fifth force active in low density regions is screened in regions of high density by shortening the range of interaction of the field  \cite{khourya,khouryb}. A more general approach utilizes Effective Field Theory (EFT) of cosmic acceleration \cite{creminelli,bloomfield}, where recent theoretical advances have shown that there exists a large model space that recovers an accelerated expansion on large scales, while reducing to Newtonian gravity on the small scales in linear theory \cite{lombrisertaylor}. Thus, there is a need for well-defined observational tests which can distinguish the many models, including their possible covariant nature, as well as $\Lambda$CDM.

In this paper, we present a novel test of gravity on galaxy cluster scales that exploits how the Chameleon mechanism modifies the gravitational potential of clusters of different masses. Specifically, MG deepens the potential in the outskirts of low mass galaxy clusters with respect to GR, but leaves the potential of high mass clusters relatively unaffected.  By taking the average ratio between the gravitational potential of high mass and low mass galaxy clusters, we show that one can unambiguously discern between Chameleon-like modified gravity theories and GR.

We note that our proposed test is complementary to probes of gravity on larger scales (1-10 Mpc) \cite{lam,hellwing,zu,xu} and can also act as a powerful cross-check of tests in galaxy cluster scales (0.1-1$h^{-1}$Mpc) \cite{SchmidtVikhHu,gRedshift,coma,wilcox,cataneo,lombriserClusterDensity,shapes,schmidt}. However, our test distinguishes itself in that it directly probes the gravitational potential and allows for a simple and elegant incorporation of theoretical predictions. 

We carry out our proposed test of gravity using synthetic dark matter halos of cosmological N-body simulations and test its viability by comparing results with analytic theory. We then utilize simulated galaxy catalogs to incorporate realistic observational systematics, including projection effects. After vetting our theoretical predictions with simulations, we show that our probe has the potential to deliver more competitive cluster-scale constraints on Chameleon \textit{f(R)} MGs than at present.

The paper is organized as follows: in Sec. II we review the Hu-Sawicki $f(R)$ gravity model and derive the theoretical expectations of our observable in both MG and GR.  In Sec. III we describe how we obtain our observable from the phase space of galaxy clusters, and detail how we do this in N-body simulations. A brief description of the N-body simulations we used is also provided. Sec. IV is devoted to putting together both our theoretical expectations and observables. More specifically, we describe how our test probes gravity in the scale of galaxy clusters.  In Sec. V we address both theoretical and observational systematics, whether or not they are significant as well as how we fold them into our final analysis.  Finally, in Sec. VI we discuss both how our probe can set competitive constraints on MG and, more generally, can also act as a powerful test of GR in the scale of galaxy clusters. We conclude in Sec. VII with some remarks on how our probe will leverage the observational capacities of future large-scale photometric and spectroscopic surveys.

 \section{Theoretical expectations}
 \subsection{Hu-Sawicki $f(R)$ gravity}
In what follows we focus on a particular model of MG: Chameleon Hu-Sawicki \textit{f(R)} gravity  ~\cite{HuSawicki}. In \textit{f(R)}  gravity, the Einstein-Hilbert action is augmented by a free function of the Ricci scalar (\textit{R+f(R)}).  This modification introduces an additional degree of freedom which can be recast as a non-minimally coupled scalar field, $f_R \equiv  \frac{df(R)}{dR}$, dubbed the \textit{scalaron}. The functional form of $f(R)$ in the Hu-Sawicki model is as follows: 

\begin{equation}
f(R) = -\bar{m}^2 \frac{c_1 (R/\bar{m}^2)^n }{c_2 (R/\bar{m}^2)^n + 1}.
\end{equation}

The $\bar{m}$ parameter sets the mass scale and is given by $\bar{m}^2= \frac{8\pi G}{3} \rho_{m0}$, where  $\rho_{m0}$ is the average density today. As such, this model is determined by three dimensionless free parameters: $c_1, c_2$ and $n$. The specific values of these free parameters can be narrowed down to those which produce expansion histories that are consistent with current cosmological constraints. In particular, these three parameters are related to the background value of the scalaron today, $f_{R0}$. For example, with $\Omega_M = 0.24$ and $\Omega_{\Lambda}= 0.76 $, we have that $f_{R0} \approx n c_1/c_2^2/(41)^{n+1}$ \cite{HuSawicki}. From now on, we parametrize these free parameters in terms of $f_{R0}$ and consider models that are phenomenologically viable. We refer the reader to Hu and Sawicki (2007) for the particularities of the model and the details of the calculations shown above.  We fix $n=1$ and consider only models with background field values in the present epoch of $|\overline{f_{R0}}|= 10^{-5}  \text{ and } 10^{-6} $, which will be denoted from now on as FR5 and FR6 respectively. Note that in our definition of $|\overline{f_{R0}}|$ we have set the the speed of light to unity. 
 
 \subsection{Gravitational potential}
 Naturally, as a coupled scalar field, the effect of the scalaron is to mediate an additional fifth force between massive bodies. Thus, the gravitational potential which massive particles experience is no longer the usual Newtonian dynamical potential of the Poisson equation ($\phi_{GR}$), but rather~\cite{lombriserClusters}:
\begin{equation}
\phi(r) = \phi_{GR}(r) - \frac{1}{2} \delta f_{R}(r).
\end{equation}
Here the  $\delta$ signifies that the background has been subtracted from the scalaron field: $\delta f_{R} = f_{R} - \overline{f_{R}}$. 

The additional scalar field  can be shielded in Chameleon Hu-Sawicki $f(R)$ gravity in order to recover the predictions made by GR in high density regions. In the high density ``Chameleon regime,'' the screening mechanism ensures we recover GR  by making $f_R \rightarrow 0$.  As such, in this regime the scalaron's field value is:
\begin{equation}
\delta f_{R}^{cham} =  -\overline{f_{R0}}.
\end{equation}
The scalar field is constant in high density regions (e.g. in the core of galaxy clusters) and can mediate no additional forces. Outside of the high density core, the field can propagate and mediate a fifth force. The range of this fifth force is determined by Compton wavelength of the field, or the inverse mass  of the scalar field  $\lambda_c \equiv m^{-1}$, which at the background and for $z=0$ is: $ m^{-1} = 32 \sqrt{\overline{f_{R0}}/10^{-4}} \text{ Mpc } h^{-1}$ ~\cite{SchmidtVikhHu}. In this ``linear regime,'' the scalaron field is given by \cite{lombriserClusters}:
\begin{multline}
\delta f_{R }^{lin}(r) = -\frac{1}{3} g(c) GM_{200}  \big\{ \Gamma(0,m(r+r_s)) e^{2m(r+r_s)} \\ +\Gamma(0,-m(r+r_s))- \Gamma(0,-mr_s) \\ - e^{2mr_s} \Gamma(0,mr_s)  \big\} \frac{e^{-m(r+cr_s)}}{r}.
\end{multline}
The upper incomplete gamma function $\Gamma(s,r)$ and $g(c)$ are given by $\Gamma(s,r) = \int_r^{\infty} t^{s-1} e^{-t} dt $ and $g(c) = [\ln(1+c) + c/(1+c )]^{-1}$. We have also used the definition: $r_s \equiv R_{200}/c $. Eq. 3 is specific to the case in which the scalaron is propagating in an Navarro-Frenk-White (NFW) density field of a galaxy cluster with concentration $c$, mass $M_{200}$, and radius $R_{200}$. The subscript 200 implies that the mass and radii are defined to be where  the density of the halo equals $\Delta_{200} = 200$ times the critical density of the universe: $M_{200}= \frac{4\pi}{3} R_{200}^3 \Delta_{200} \rho_{crit}$.    In practice, the concentration is an NFW fitting parameter attained by fitting the cumulative mass profile \cite{mamon,NFW}:
\begin{equation}
M(<r) = g(c) M_{200}   \bigg[ \ln\bigg(1+ \frac{r}{r_s} \bigg) - \frac{r}{ r+r_s}\bigg].
\end{equation}

The transition from the Chameleon (Eq. 3) and Linear regimes (Eq. 4) is efficient, so we model it as being instantaneous and match them using,
\begin{equation}
\delta f_{R} = \min (\delta f_{R}^{lin}, \delta f_{R}^{cham}).
\end{equation}
As shown by ~\cite{lombriserClusters}, this approach agrees with the numerical solution to the scalaron equation of motion in the vicinity of an NFW density field. Note that the theoretical uncertainty is negligible when compared to observational uncertainties (thoroughly explained in the Sec. V below). 

The GR potential ($\phi_{GR}$) is the usual gravitational potential that satisfies the Poisson equation and therefore determines the motion of massive particles. $\phi_{GR}$ can be attained by solving the Poisson equation with the NFW mass profile of Eq. 4,

\begin{equation}
\phi_{GR}(r) =  -g(c) \ln\bigg( \frac{r}{r_s}+1\bigg) \frac{GM_{200}}{r} .
\label{eq:phi_gr}
\end{equation}

\subsection{Escaping a galaxy cluster in \\ an expanding universe}

However, our observable, as explained in the next section, is the escape velocity profile ($v_{esc}(r)$) which is related to the potential set by both the gravity of the cluster and the expanding universe ($\Phi(r)$). This effective potential relates to the escape velocity profile as usual: $-2\Phi(r) = v^2_{esc}(r) $. Following \cite{behroozi}, we derive $v_{esc}(r)$ for a cluster described by an NFW density in an expanding universe with GR as its prescription for gravity,
\begin{eqnarray}
-2\Phi_{GR}(r) &=& v^2_{esc,GR}(r),  \\
               &=& -2[\phi_{GR}(r) - \phi_{GR}(r_{eq})] - qH^2 [r^2-r_{eq}^2] \nonumber.
\label{eq:phi_gr_big}
\end{eqnarray}
$r_{eq}$ is the ``equivalence radius'', defined to be the point at which the acceleration due to the gravitational potential of the cluster and the expanding universe are equivalent: $r_{eq}\equiv(\frac{GM_{200}}{-qH^2})^{1/3}.$ $q$ is the deceleration parameter and $H$ is the Hubble parameter. Using Eq. 1, in an expanding Chameleon \textit{f(R)} gravity universe instead we have:
\begin{eqnarray}
-2\Phi_{f(R)}(r) &=& v_{esc,f(R)}^2(r), \\
               &=&  v_{esc,GR}^2(r)+ \big[\delta f_R(r)- \delta f_R(r_{eq})\big] \nonumber.
\label{eq:vesc_fr}
\end{eqnarray}
Eqs. 7 and 8 use NFW parameters that have been measured from the cluster density profiles. In real data, these would be measured using the observed weak lensing shear profile around clusters which is unaffected by the effects of $f(R)$ gravity (see the appendix of \cite{springer}). Our test is then built to compare the matter-inferred GR or \textit{f(R)} gravity potential profiles to the observed dynamical escape velocity profile.  

\section{Observables}
We measure the escape velocity of galaxy clusters through the technique (developed in \cite{diaferio}), in which the Newtonian gravitational potential profile is reconstructed from the escape velocity ``edge'', identified from the cluster radius-velocity space (i.e. the ``phase space''). This is a well-developed and well-tested technique and it has been used in numerous studies of the mass profiles of galaxy clusters \cite{dan1, serra, lemze, rines, andreon, geller}. The escape edge for a given halo is constructed by taking the maximum velocity in the particle phase space for each $r/R_{200}$ bin in 0.05  intervals. In GR simulations, the observed escape edge has been shown to recover the theoretical $v_{esc}$ to $\sim 5\%$ accuracy, depending on the model used \cite{miller}. 

\subsection{N-body simulations}
To construct our potential profiles from the escape velocity edge we utilize the particles in N-body simulations developed in ~\cite{sims1,sims2} which have identical initial conditions, but incorporate either GR or \textit{f(R)} as their prescriptions for gravity. The cosmological parameters used for these simulations are as follows: $\Omega_m=0.24, \Omega_{\Lambda}=0.76, h=0.719, n_s=0.961$ \cite{sims2}. Here our focus is on the simulated  GR, FR6, and FR5 runs. Briefly, the adaptive mesh refinement N-Body simulations take place in a 1.5 Gpc $h^{-1}$ cube with 1024$^3$ resolution and a particle resolution of $6.2 \times 10^{10} M_{\odot} h^{-1} $. The large box size is required to provide the number of high mass clusters required in our analysis. The halos are defined by the Amiga Halo Finder or AHF \cite{AHF}. 

\section{Probing gravity}  

 \begin{figure*}
\centering
  \includegraphics[width=1\textwidth]{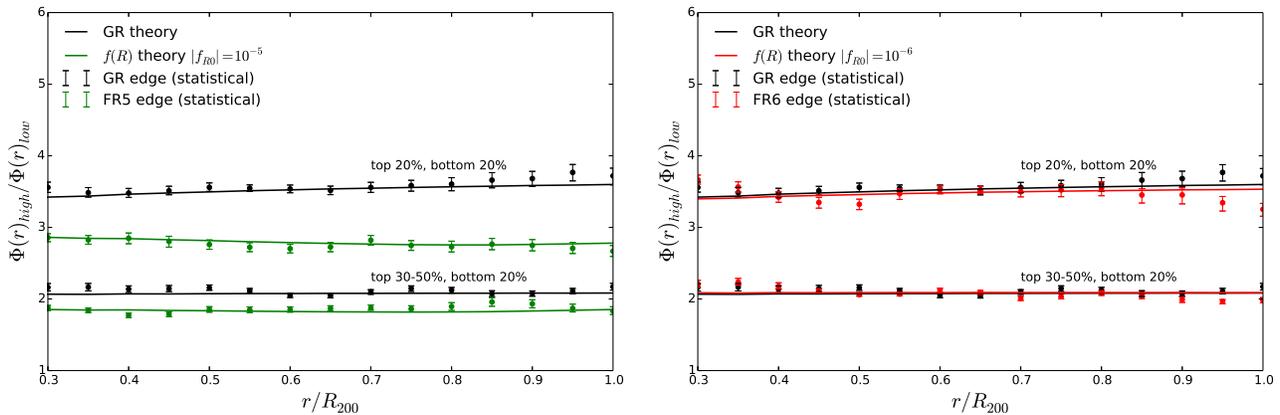}
  \caption{The $z=0$ gravitational potential ratio between high and low mass bins of synthetic galaxy clusters for the GR (black), the FR5 (\textit{left}) parametrization of $f(R)$ gravity (green), and the FR6 (\textit{right}) parametrization (red). The points are the average of the square of the observed escape velocities for each bin in radius and mass.  The errors are 1$\sigma$ on the mean from boot-strap re-sampling. The solid lines represent the theoretical predictions using the NFW density parameter (Eqns. 7-9). Note the  $\gtrsim 5 \sigma$ level difference between the GR and FR5 ratios. The percentages denote how we separate the high and low mass bins. Note that the separation between GR and \textit{f(R)} potential ratios increases with increasing separation in the mass bins.}
\label{fig:potential_ratio}
\end{figure*}

To differentiate between GR and MG we probe the \textit{ratio} of the averaged gravitational profile of high mass clusters, to the averaged gravitational potential of low mass clusters. We infer the ratio from N-body simulations through the aforementioned technique and compare our results to the aforementioned theoretical expectations for each respective theory of gravity. More specifically, our probe is encapsulated in the following equation: 

\begin{equation}
\frac{\Phi_{high}(r)}{ \Phi_{low}(r)}  \equiv \frac{\langle{ v_{esc,high}^2(r)}\rangle}{ \langle v_{esc,low}^2(r) \rangle}.
\label{eq:phiratio}
\end{equation}
Why separate clusters into two mass bins? The reasons are twofold. First, as shown in \cite{KoyamaMASS} the Chameleon mechanism induces a mass dependent screening effect which leads to high mass clusters being screened and low mass clusters being increasingly unscreened. Relative to their high mass counterparts, less screened low mass clusters will exhibit higher escape velocity edges, resulting in a reduced potential ratio compared to expectations from GR. The cluster potential ratio is a smoking gun test of modified theories of gravity employing the Chameleon screening mechanism. Secondly, the ratio allows us to undercut both observational and theoretical systematics. Precisely how this ratio allows us to do so is thoroughly explained in the next section.

Now, to compare clusters in each of the three (GR, FR6, and FR5) simulations we employ one-to-one matching. We find cluster-sized over-densities at the same positions at $z=0$ across all three simulations using a halo's center positions from an AHF-generated halo catalog \cite{AHF}. We begin with 100 halos uniformly sampled in log mass between $\sim 10^{14}- 10^{15}$M$_{\odot} h^{-1}$. The 100 halos per simulation are then binned into a \textit{low mass bin} and a \textit{high mass bin} each of which corresponds to selecting a percentile of the of least and most massive clusters. The specific mass bin ranges are as follows. For GR, the low mass bin is: $9.10\times10^{13} - 1.96\times10^{14} M_{\odot}  h^{-1}$  and the high mass bin is: $7.48\times10^{14} - 1.58\times10^{15} M_{\odot} h^{-1}$. For FR6, the low mass bin is: $9.13\times10^{13} - 1.97\times10^{14} M_{\odot}  h^{-1}$  and high mass bin is: $7.34\times10^{14} - 1.58\times10^{15} M_{\odot} h^{-1}$. For FR5, the low mass bin is: $1.16\times10^{14} - 1.94\times10^{14} M_{\odot}  h^{-1}$  and the high mass bin is: $7.49\times10^{14} - 1.58\times10^{15} M_{\odot} h^{-1}$. 

For each of the simulation halos, we attain both the phase space escape velocity profile through the aforementioned technique, as well as the potential profile based on matter density NFW fit (Eq. 5). With the former we can construct the ``observed'' dynamical gravitational potential profile and with the latter the prediction from GR or \textit{f(R)} gravity. We then take the ratio between the averaged high mass edge profiles and the averaged low mass edge profiles (Eq. 10).

The resulting averaged profile ratios are shown in Figure 1. The errors are 1$\sigma$ on the mean from boot-strap re-sampling with replacement. The solid lines represent the theoretical predictions using the NFW density parameter (Eqns. 7-9). Note that we chose to present the ratios as a function of $r/R_{200}$ rather than $r$ as a way to remove the side effects that arise when comparing clusters of different masses. When plotted as a function of $r$ the potential ratio of Figure 1 attains a positive slope which arises as a result of different underlying mass profiles. Thus, re-scaling by each cluster's respective $R_{200}$ flattens the profile and makes the potential ratios of Figure 1 a much cleaner and clearer observable: one that depends on the amplitude of the potential and not the shape of the potential profile.

As Figure 1 also shows, our theoretical predictions can successfully reproduce our (simulated) observable to high precision. On that figure we also plot the averaged potential ratio for different mass bins. One important conclusion from this is that the closer we bring the two mass bins, the more attenuated the difference between GR and $f(R)$ becomes. We would like to mention that dynamical differences in Fig. \ref{fig:potential_ratio} are solely due to modifications to gravity and not to mass differences in the simulation. Comparing the average mass ratios: $\langle M_{200,high}\rangle/ \langle M_{200,low} \rangle = 6.40, 6.39, 6.02$ for GR, FR6 and FR5 respectively, we conclude that averaged mass ratio differences are negligible when compared to the $\sim 25\%$ difference between the GR or FR6 and the FR5 potential ratios shown in Fig. \ref{fig:potential_ratio}.

Given our ability to precisely predict our observable (as demonstrated by Figure 1), we demonstrate the theoretical lower limit on how well our new probe can constrain Chameleon $f(R)$ gravity. Cosmic variance is the largest component of the uncertainty on the potential ratios for the small samples we examine. We study a more realistic synthetic data-set using the 2-dimensional projected phase spaces in the following sections.

 
\section{Systematics}
In order to assess the viability of our probe, we carefully consider relevant systematics. More specifically, we focus on: the systematic error induced by cosmic variance, statistical errors that arise from projection effects, as well as additional systematics that arise from sampling of galaxies in clusters.

One relevant systematic arises from the fact that we are using a very small sample of the clusters in the Universe to generate Figures 1 and 2. We chose this sample size as it reflects the scale of data we have today for measuring both weak-lensing profiles and also with significant spectroscopic follow-up \cite{Rines,Geller13,hoekstra12}.

The dashed lines band as shown in Figure 2 incorporates the systematic uncertainty due to cosmic variance on a sample of this size (20 clusters per mass bin) which was measured to be 10\% using a larger set of GR simulations (the Millennium simulations  \cite{Springel}). In other words, the observed ratio $\frac{\Phi_{high}(r)}{ \Phi_{low}(r)}$ can vary as much as 10\% when measured for a sample size as small as $\sim$ 20 clusters per mass bin. This is an important systematic which decreases as the sample size increases.

We also assess environmental screening, in which some fraction of the lower mass clusters in the $f(R)$ simulations could be screened due to large-scale over-densities \cite{zhao}. If this were to happen, the observed FR5 and FR6 ratios would be higher (closer to GR) than predicted by the theory. We control for random noise due to variations in the large-scale environments of the clusters by using one-to-one matching for the halos across the GR, FR5 and FR6 simulations. Also, we ensure that none of the clusters lie within the virial radii of other clusters. Finally, we can evaluate our dataset for possible effects from environmental screening using Figure \ref{fig:potential_ratio}, where the FR5 and FR6 predictions match our measurements to high precision. If screening were present, one would find the FR5 and FR6 measurements of the $f(R)$ simulation data to be closer to the GR simulation data. We note that the absolute accuracy of the cluster sample used in Figure \ref{fig:potential_ratio} is limited by cosmic variance of order 10\%. We conclude that the effects of screening on this test are smaller than systematic and statistical uncertainties on our measurements, so long as non-merging clusters are chosen for the analysis.

 \begin{figure}
  \centering
  \includegraphics[width=1\linewidth]{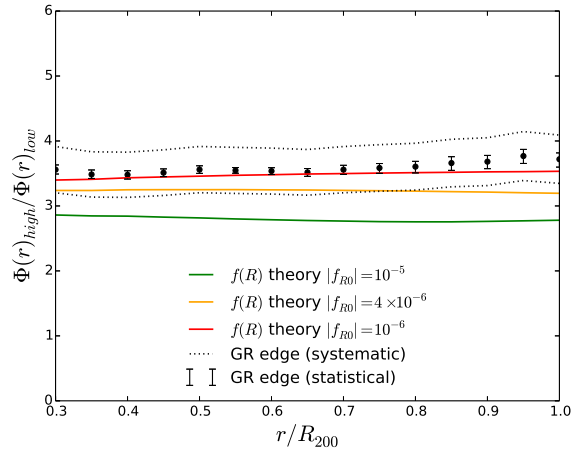}
  \caption{The simulated and theoretical averaged GR and $f(R)$ gravity potential ratios for the same clusters of Figure 1 (top 20\% and bottom 20\% mass bins). We have added the theoretical prediction for  $|f_{R0}| = 4 \times 10^{-6}$ gravity (yellow) and include not only the GR statistical error on the mean (black error bars) but also the 10\% systematic error due to cosmic variance (black dashed lines). This figure demonstrates how we would detect and/or constrain MG. Assuming GR as our ``observation,'' the data is contained within the  black dashed bands. We can therefore ask the question: which of the three plotted models best describes the data? We conclude our probe can, including systematics, successfully discern between GR and $|f_{R0}| = 4 \times 10^{-6}$ at $1\sigma$.Note that this result is attained with only 40 clusters (20 in each mass bin) and the potential is inferred from 3-dimensional phase space synthetic data. We reproduce this plot for a projected DESI-like data-set in Fig. 4.}
\end{figure}


We use the Millennium simulations with the light cone data provided by \cite{Henriques}  to investigate projection effects, which is likely the dominant component of the error on the potential ratio. To carry out our proposed test with physical (rather than synthetic) data, the escape velocity profile of a cluster would be inferred from the line-of-sight velocities of galaxies in that given cluster, rather than the radial component of the 3D velocity. Fortunately we can transform, roughly, the line-of-sight velocities into 3D velocities and vice versa. As shown in \cite{DiaferioProject} the mapping between the radial 3D escape velocity profile considered above ($v_{esc}(r)$) and the line-of-sight escape velocity profile inferred from data ($v_{esc,los}(r)$) is as follows:
 \begin{equation}
v^2_{esc,los}(r)= g^{-1}(\beta) \times v^2_{esc}(r)  \\
\label{eq:projected}
\end{equation}
Where $g(\beta)$ is given by $g(\beta) =3-2\beta(r)/1- \beta(r)$ and is of $\mathcal{O}(1)$. $\beta(r)$ is given by 
$\beta(r) = 1- \langle v_{\theta}^2 + v_{\phi}^2 \rangle/  2\langle  v_{r}^2\rangle$ \cite{DiaferioProject}. As such, in projected space Eq. 10 is instead,
\begin{equation}
\frac{\Phi_{high,los}(r)}{ \Phi_{low,los}(r)}  \equiv\frac{ \langle{g(\beta_{low})}\rangle}{\langle{g(\beta_{high})}\rangle} \frac{  \langle{ v_{esc,high}^2(r)}\rangle}{ \langle v_{esc,low}^2(r) \rangle} .
\label{eq:phiratio_proj}
\end{equation}
However, the ratio of the averaged $g(\beta)$ profiles for high and low mass systems is $\langle g(\beta_{high})\rangle / \langle g(\beta_{low})  \rangle \approx  1$ and so the line-of-sight potential ratio is the same as the 3D potential ratio:
\begin{equation}
\frac{\Phi_{high,los}(r)}{ \Phi_{low,los}(r)} \approx \frac{\Phi_{high}(r)}{ \Phi_{low}(r)}
\label{eq:phiratio2}
\end{equation}
Therefore, our expectation is that by dividing out the averaged cluster potential profiles we eliminate the necessity to estimate the anisotropy profile. We note that another observational challenge lies in eliminating line-of-sight galaxies that may not be cluster members and will therefore contaminate our phase space. 

We recognize that there exist few observational surveys containing clusters around the mean mass in our low-mass subsample.  However, with new imaging and spectroscopic surveys, we expect future datasets to provide excellent weak lensing mass profiles as well as significant spectroscopic follow-up. 

Therefore, we consider two cases of projection. The first uses an ensemble cluster dataset, where the weak-lensing mass profiles and the spectroscopic potential profiles are inferred from averaged or stacked datasets as would be measured using current facilities. The second uses a much larger sample of clusters based on deeper data, where the weak-lensing and spectroscopic potential profiles could be measured individually for the clusters and then averaged.

\begin{figure}
  \centering
  \includegraphics[width=1\linewidth]{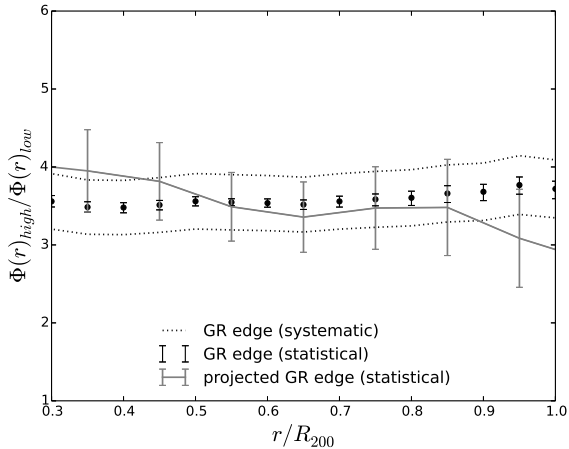}
  \caption{The gravitational potential ratio of Fig. 1 (black dots with statistical error) and the gravitational ratio as inferred from an ensemble of clusters (gray line with statistical error) from the light cone data of \cite{Henriques}. The high and low mass cluster ensembles are made up of 5 different ensembles each with 10 clusters that include 500 galaxies each. The error bars are $1\sigma$ error on the mean from boot-strap re-sampling. This nearly flat transfer function incorporates numerous observational systematics when going from 3D to realistic observational data.}\label{fig3}
\end{figure}

\subsection{Stacked cluster ensembles}
 To build a cluster ensemble we superimpose the phase-spaces of individual clusters. In particular, we use 500 galaxies per phase space with 10 high mass and 10 low mass clusters to create a high and a low mass cluster ensemble. The masses of the clusters are chosen to match the average masses of the sample used in Figures 1 and 2. The galaxies used to populate the phase spaces are all brighter than an r-band magnitude of 17.7 and the clusters are within z=0.15 such that this is an SDSS-like stacked ensemble of clusters. As before, we compute the averaged potential ratios (Eq. 10) and the boot-strapped error bars. The result is shown on Fig.\ref{fig3} (solid gray line). The scatter dots with statistical error represent the 3D inferred edge (as in Fig. 1 and 2). We find that over the range $0.4 \le r/R_{200} \le 0.9 $, the 2D projected and the 3D ratios are statistically identical. At the same time, our constraints on $|f_{R0}|$ are robust to small (5\%) corrections in the ratio as we go from 3D to 2D. This confirms the expectation detailed above which speculated that the averaged ratio of potential profiles allows us to undercut complications that arise from working with projected data. However, Fig. 3 also shows that the errors from projection are as large as cosmic variance for a small sample of $\sim$20 clusters. 

\subsection{DESI Bright Galaxy Sample forecast}
In order to decrease the errors, we investigate a much larger (by a factor of $\sim 10$) sample of clusters. Within the Henriques light cone  \cite{Henriques} we generate a Dark Energy Spectroscopic Instrument (DESI) Bright Galaxy Sample-like selection function. While for the previous studies we constrained the observed galaxy population to brighter than the SDSS main sample, we now use galaxies to an r-band magnitude limit of 19.1. We still focus on clusters within z=0.15. The deeper magnitude limit increases the number of clusters with greater than 50 galaxies in the phase space to many hundreds per mass bin. This sample is an approximation to what may be observed with DESI and would allow us to probe both a wider and deeper sample of the sky, thereby undercutting the effects of cosmic variance and increasing the total number of galaxy clusters in our sample.

We re-create the mass bins chosen in the previous sample (see section IV) by choosing clusters with masses between $ 10^{14}/h < M_{200}/M{\odot} < 2.1\times 10^{14}/h $ for the low mass sample, and $ M_{200}/M{\odot} >  6\times10^{14}/h $ for the high mass sample. The average mass for the low and high mass sample respectively are: $1.4 \times 10^{14} M{\odot} / h$ and $8.2\times 10^{14} M{\odot} / h$. From those two mass bins we then make a conservative cut by picking clusters that contain at least 50 galaxies between $|v_{los}| \leq 2000 \text{ km/s}$ within 3 Mpc from the cluster center. 

To predict the potentials in this sample, we use the halo masses of each cluster and derive concentrations from the mass-concentration relation provided in \cite{Duffy08}: $c_{200}(M_{200}, z) = A_{200} (M_{200}/M_{pivot})^{B_{200}} (1+z)^{C_{200}}$. Where $A_{200}= 5.71$, $B_{200} = -0.084$,  $C_{200} = -0.47$ and $M_{pivot} = 2\times 10^{12}  M_{\odot}h^{-1}$. This sample assumes that weak lensing masses are unbiased. The width of our mass bins are larger than the one sigma mass scatter in the weak lensing observable \cite{bandk}. Because of the wide width of our mass bins, we can ignore weak lensing mass uncertainties when calculating the theoretical predictions.

In Figure 4, we show the averaged potential ratio between the two mass bins (Eq. 10) with both statistical (boot-strapped as before) and systematic error (cosmic variance) for the aforementioned DESI-like sample. As before, we also show the theoretical predictions for both GR and $f(R)$ gravity. Several conclusions regarding systematics affecting our probe may be drawn from Figure 4:

\begin{itemize}

\item Our GR 3D theory (solid black line) can accurately predict the potential ratio generated with projected synthetic data. This confirms that we can successfully divide out projection effects by taking the ratio of averaged potentials (as implied by Eqns. 12 and 13). 

\item Projection effects increase the statistical error relative to the unrealistic 3D ratio. This is expected and discussed in \cite{dan2}. In particular, compare the few percent statistical error on the 3D ratio of Figures 1 and 2 with the $\sim 8\%$ statistical error of Fig. 4. However, while projection increases the statistical error on the ratio, the cosmic variance of a larger sample goes down as the square root of the fractional increase in the sample size, which for a DESI-like sample is $\sim$10 times as many clusters as used in our previous results (Figures 1, 2, and 3.)

\item The phase spaces only need to be moderately populated with 50 galaxies between $|v_{los}| \leq 2000 \text{ km/s}$ within 3 Mpc from the cluster center. This is necessary in order to measure the escape velocity edge. By selecting the 50 brightest galaxies to create the phase spaces, the test is immune to color bias, at least to the level probed by the simulated galaxy catalog. We also find that if we under-populate the phase space, we can bias our low mass cluster potential profiles. This is a known effect and studied in detail in \cite{dan1}.
\end{itemize}

We also investigate whether the assumed mass-concentration relation affects our potential ratios. We examine a range of uncertainties in the parameters which describe the relation and find that the potential ratio profiles vary less than 1\%. This is because all mass-concentrations relations are quite flat at the cluster masses we study here and also because the mass difference between the high and low mass sub-samples is quite small.

Generally, we note that the utilization of the ratio of potentials mitigates systematic effects of the observables and theory beyond the aforementioned projection effects. For instance, \cite{miller} found that NFW density profiles predict NFW potentials that are biased high by 10-20\%. Einasto profiles on the other hand show $<$5\% biases. However, they also showed that there is no difference between the predicted and observed phase space escape velocities as a function of halo mass. We tested that by comparing the GR prediction from Eq. \ref{eq:phiratio} using NFW density fits to the more accurate Einasto density profile fits. We find that the ratios agree to within a percent. In other words, while the NFW mass profile systematically overestimates the cluster potential profiles, the ratio of NFW potentials is unbiased. Similar arguments can be made about other possible systematics, including line-of-sight effects, velocity bias, and velocity anisotropies as mentioned in the bullet points above \cite{sven15,dan1,lemze}.

\section{Statistical Constraints on $f(R)$  gravity} 

\begin{figure}
  \centering
  \includegraphics[width=1\linewidth]{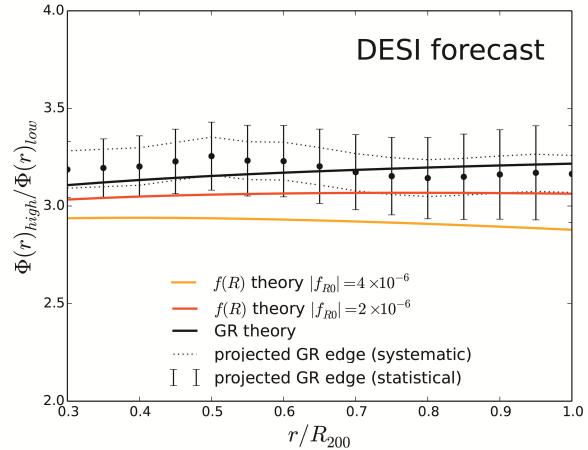}
  \caption{The projected gravitational potential ratio for a DESI-like galaxy cluster sample (black dots and bootstrap error on the mean) as inferred from synthetic galaxy clusters from the light cone provided by \cite{Henriques}. Compared to Fig. 2 (where the ratio is inferred from a 3D phase space) we see that projection significantly increases the statistical error. However, the DESI sample of Fig. 4 is significantly larger as it contains 9.6 times many more clusters and so the systematic error is significantly reduced
  (286 low mass clusters and 96 high mass clusters). We can determine which gravitational theory best matches the "observation." To visualize this, we plot both the GR theoretical prediction (black line) and two $f(R)$ theoretical predictions (orange-red and yellow). We conclude that our probe can differentiate between GR and $|\overline{f_{R0}}|= 4 \times 10^{-6}$ ($2 \times 10^{-6}$) at $>5\sigma$  ($>2\sigma$).}\label{fig4}
\end{figure}

 Given our analysis of systematics detailed above, we can provide a robust estimate of how well our probe will be able to constrain MG on the two different cluster samples we simulate. 
 To differentiate between GR and $f(R)$ gravity, we assume that we live in a GR universe with a $\Lambda$CDM cosmology (i.e. our  ``measurements'' are the black dots of Figure 1-4), and calculate the $\chi^2$ between the measurement and the theoretical predictions.  The statistical GR error is calculated from boot-strap re-sampling of the mean and the systematic GR error takes into account the cosmic variance.  The uncertainty is taken to be the combined statistical and systematic errors added in quadrature.

 We first compare against a stacked ensemble of clusters. We assume an SDSS-like cluster sample with 10 stacked clusters per ensemble and 5 cluster ensembles per mass bin. Recall that individually, the clusters are too poorly sampled in their spectroscopy. However, the two ensemble clusters would have a high signal-to-noise weak lensing mass profile as well as a well-determined escape velocity profile. In this case, the cosmic variance term dominates the error budget. The $\chi^2$ test indicates that we can differentiate GR from $|\overline{f_{R0}}|= 5.5\times 10^{-6}$ at $2\sigma$. 

 By increasing the sample size we can improve these constraints. In particular, see Fig. 4, where we show that with the DESI-like sample of galaxy clusters, we can constrain $|\overline{f_{R0}}| = 4 \times 10^{-6}$ ($2 \times 10^{-6}$) at $>5\sigma$  ($>2\sigma$). 
 
 These are competitive forecasts compared with the two most recent galaxy cluster-scale constraints. For example, \cite{wilcox} presents an analysis of 58 clusters in the XMM sample that constrains  $|\overline{f_{R0}}| < 6\times10^{-5}$ at 95\% confidence. Similar to the potential ratio test presented here, \cite{wilcox} focuses only on cluster scales, i.e., within the virialized region of clusters. Using stacking on only 50 clusters per mass bin and after including both statistical and systematic uncertainties on the observable, our potential ratio test can achieve an order-of-magnitude improvement over observational constraints set by \cite{wilcox}. The most competitive cluster-scale constraint is set by \cite{cataneo} with a cluster abundance analysis. They find that  $|\overline{f_{R0}}| < 1.62\times10^{-5}$ at 95\% CL. Our test should be able to differentiate $|\overline{f_{R0}}| = 1.62\times10^{-5}$ and GR at $>>5\sigma$.

\section{Conclusions} 
We propose a new test for gravity within galaxy clusters that leverages the ways in which  modifications to gravity alter the dynamical potential while leaving the weak-lensing inferred potential profile unchanged. We take the ratio of the squared escape velocity profile for high and low mass clusters as our observable. We do this for two reasons: 1) it leverages the fact that Chameleon screening leaves the dynamics of the high mass clusters unaffected compared to the low mass clusters and 2) it removes any systematic that is present in both samples, such as velocity bias and velocity anisotropy. While this test can be applied generally to any new model for gravity which has this property, we test it against Chameleon $f(R)$ gravity.

We first use simulations to show that particles, as tracers of the dynamical gravitational potential within galaxy clusters, do have enhanced escape velocity profiles compared to expectations from their non-dynamical (i.e. particle) masses. We then use mock galaxy catalogs to understand the role of systematics in quantifying how well we can rule out $f(R)$ gravity. We study two cases. The first utilizes a realistic but rather small number of clusters which we stack to create 2 cluster ensembles with different mass bins. We also study a second dataset which is much larger and more representative of what we expect from future surveys. In the former, cosmic variance dominates the systematic error budget, while in the latter 2D projection effects dominate. In either case, we find that our probe is more sensitive, by an order of magnitude, over current cluster-based tests for Chameleon $f(R)$ gravity.  More specifically, we have quantified our prediction for a DESI Bright Galaxy Sample-like set of clusters to push down current constraints to $|\overline{f_{R0}}| < 2 \times 10^{-6}$ at $>2\sigma$.

The test of gravity that we propose here is designed to leverage the next generation large-scale photometric and spectroscopic surveys (e.g., the Dark Energy Survey \cite{Diehl} and the Dark Energy Spectroscopic Instrument--DESI) providing high quality weak-lensing mass profiles of clusters and deep and plentiful spectroscopic follow-up of the cluster phase spaces. While we focus on scales $r < R_{200}$, we note that the difference between the GR and $f(R)$ gravity ratio increases as we go out to larger radii. This is due to the scalaron's fifth force being more effective in the outskirts of clusters where the density is lower. We also note that the difference between the GR and $f(R)$ gravity decreases with increasing redshift. We have not yet included these properties into our statistical quantification, but we expect that they will prove useful in setting even tighter constraints. We will explore these in a future effort.

\section{Acknowledgements} The authors are grateful to Elise Jennings for insightful and clarifying discussions. This material is based upon work supported by the National Science Foundation under Grant No. 1311820 and 1256260. GBZ is supported by the Strategic Priority Research Program "The Emergence of Cosmological Structures" of the Chinese Academy of Sciences Grant No. XDB09000000. KK is supported by the UK Science and Technology Facilities Council grants ST/K00090/1 and the European Research Council grant through 646702 (CosTesGrav).

\end{document}

%% file: author_list.tex
%

%
\author{Alejo Stark}
\affiliation{Department of Astronomy, University of Michigan, Ann Arbor, MI 48109 USA}
\author{Christopher J. Miller} 
\affiliation{Department of Astronomy, University of Michigan, Ann Arbor, MI 48109 USA}
\affiliation{Department of Physics, University of Michigan, Ann Arbor, MI 48109, USA}
\author{Nicholas Kern}
\affiliation{Department of Astronomy, University of California, Berkeley, CA 94720 USA}
\author{Daniel Gifford}
\affiliation{Department of Astronomy, University of Michigan, Ann Arbor, MI 48109 USA}
\author{Gong-Bo Zhao}
\affiliation{National Astronomy Observatories, Chinese Academy of Science, Beijing, 100012, P.R.China}
\affiliation{Institute for Cosmology and Gravitation, University of Portmouth, Portsmouth, PO1 3FX, UK}
\author{Baojiu Li}
\affiliation{Institute for Computational Cosmology, Durham University, Durham, DH1 3LE, UK}
\author{Kazuya Koyama}
\affiliation{Institute for Cosmology and Gravitation, University of Portmouth, Portsmouth, PO1 3FX, UK}
\author{Robert C. Nichol}
\affiliation{Institute for Cosmology and Gravitation, University of Portmouth, Portsmouth, PO1 3FX, UK}

%
%

%
\vskip 0.25cm